\documentclass[12pt]{article}
\usepackage{epsfig}
\begin{document}
\newcommand{\be}{\begin{equation}}
\newcommand{\ee}{\end{equation}}
\begin{center}
                 {\large {\bf  Ring-like events:\\
               Cherenkov gluons or Mach waves?}}
\vspace{0.5cm}

{\bf I.M. Dremin}\\
\vspace{2mm}

{\it Lebedev Physical Institute, Moscow, 119991}

\vspace{0.2cm}
\end{center}


\noindent {\it LANL ArXiv}: hep-ph/0507167

\begin{abstract}
Similar to electromagnetic forces, strong interactions might induce such 
collective effects as Cherenkov and Mach waves. Their conical structure
would be responsible for specific ring-like events. The theoretical and 
experimental arguments in favor of these phenomena are discussed and their 
most important features are described.
\end{abstract}

\section{Introduction}

Cherenkov photons and Mach shock waves are wellknown collective effects in 
physics.
Both have the similar origin - wave radiation by a body moving with speed 
exceeding the wave velocity $c_w$, i.e. the phase velocity. For Cherenkov
photons, to be emitted by a charged particle moving with the velocity $v$, 
the phase velocity of light in a medium must be less than $v$. For 
Mach waves, this is the velocity of sound. If $z$-axis is chosen along the body 
propagation, then in both cases emission in an infinite medium at rest is 
directed along the cone with the polar angle $\theta $ defined by the condition
\be
\cos \theta = \frac{c_w}{v}.     \label{cos}
\ee
For Cherenkov photons $c_w=c/n$ where $n$ is the index of refraction, for
Mach waves $c_w=c_s$ where $c_s$ is the sound velocity in the medium.
These are collective effects determined by the presence of the medium.
The wave fronts have a conical shape with an opening angle $\theta _f$
with respect to the direction of body (particle) motion given by 
$\theta _f=\frac {\pi }{2}-\theta $. Photons are emitted perpendicular
to the wave front. Rings of light in Cherenkov detectors are used to
measure $v$.
Emission of Cherenkov photons is due to the polarization of atoms induced
by a beam of charged particles. Mach shock waves are characterized by the
pressure variation. Both effects are consequences of electromagnetic 
forces acting in the medium.

{\it Are there analogues of these effects for strong interactions?}
Many indications in favor of it exist but further studies are necessary. 
Here one should rely on quantum chromodynamics (QCD) with 
quarks and gluons acting as partons in high energy interactions. Almost 
massless quarks with spin 1/2 and massless vector gluons recall electrons
and photons but color forces between them with self-interaction of
gluons lead to such new properties as asymptotic freedom and confinement.

\section{Cherenkov gluons}

The intuitive picture which comes to mind is to consider the impinging partner
in $hh, hA, AA$ collisions
as a bunch of partons passing through a hadronic medium. A target hadron or 
nucleus can be treated as a nuclear slab with a definite index of 
{\it nuclear} refraction.
The analogue to Cherenkov photons would be Cherenkov gluons emitted by a 
parton entering this hadronic medium. The notion of 
Cherenkov gluons was proposed long time ago \cite{d1, d2} and 
experimental indications in favor of this effect appearing from 
time to time are quite extensive \cite{addk, alex, masl, arat, maru, 
kit, dlln, adam, elna, agab, adk, swan, ad2, gogi, dikk, voka}.

\subsection{The nuclear index of refraction}

The necessary condition for this effect is the excess of the real part of
the index of refraction over 1. In electrodynamics, the index of refraction 
$n$ is related to the forward scattering amplitude of photons in a 
medium  $F(\omega )$ as
\be
n(\omega )=1+\Delta n =1+\frac {2\pi N}{\omega ^2}F(\omega ).   \label{nom}
\ee
Here $\omega $ is the photon frequency, $N$ is the density of the scatterers
(inhomogeneities) of the medium. The amplitude $F(\omega )$ is normalized by 
the optical theorem
\be
{\rm Im} F(\omega )= \frac {\omega }{4\pi }\sigma (\omega ),    \label{opt}
\ee
where $\sigma (\omega )$ is the total cross section of photon interaction
in the medium.

In what follows, we use the rest system of one of the colliding hadrons or 
nuclei in analogy with classical situation of charged particles passing
through Cherenkov detectors.

If the formula (\ref{nom}) is applicable in QCD, one should use the forward 
scattering amplitude of gluons $F_g(\omega )$ in a hadronic medium as well 
as the density of partons. The latest one is related to the structure functions.
It is more difficult to say something about $F_g(\omega )$.

To provide some estimates, it was proposed \cite{d1} to come back to pre-QCD
times. At those days, nucleons and pions were considered as elementary 
entities with pions being the quanta of radiation. Namely in this way the 
ideas about nuclear Cherenkov effect were first promoted \cite{wada, igur,
bind, yeku, ceri, cgla, smrz, inic, zlom}. No discussion of the nuclear index of 
refraction was attempted at that time. It was {\it ad hoc} assumed that the
necessary condition can be somehow satisfied. The purpose was to explain 
main mechanism of particle production by this effect. It has failed.

If considered in the rest system of the target, the density of scatterers
inside a nucleus would be approximately given by the inverse volume of
a nucleon, i.e. $N\approx 3m_{\pi }^3/4\pi $ ($m_{\pi }$ is the pion 
mass). Thus
\be
{\rm Re} n(\omega )=1+\Delta n_R(\omega )=1+\frac {3m_{\pi }^3}{8\pi \omega}
\sigma (\omega )\rho (\omega ),     \label{ren}
\ee
where $\rho (\omega )={\rm Re} F/{\rm Im} F$ and now $F(\omega ) \;(\sigma (\omega ))$ 
is the pion-nucleon amplitude (cross section). The necessary 
condition\footnote{More 
precisely, as seen from Eq. (\ref{cos}) the necessary condition should read 
$\Delta n_R>1/2\gamma ^2$ with $\gamma $ being $\gamma $-factor of the
impinging partner.} is fulfilled if ${\rm Re} F>0$ (or $\rho >0$).

It is known from experiment that there are two energy regions where real 
parts of hadronic forward scattering amplitudes become positive. At low 
energies, it happens in a half of any resonance shaped as 
$F(\omega )\propto (\omega - \omega _0+i\Gamma /2)^{-1}$ so that
at $\omega >\omega _0$ one gets ${\rm Re} F(\omega )_{max}=\Gamma ^{-1}$ and 
\be
\Delta n_R^r=\frac {3m_{\pi }^3}{2\omega _r^2\Gamma}. \label{del}
\ee
Here $\omega _r$ is the pion energy required to produce a resonance. It
can be of the order of $m_{\pi }$. Since the widths $\Gamma $ are of the 
order of hundred MeV for known resonances, $\Delta n_R^r$ can be of the 
order of 1. 

At high energies, according to experimental data on $pp$ and 
$\pi p$ amplitudes, $\rho (\omega )$ also becomes positive above some
threshold, which is quite high ($\omega _{th}>70$ GeV, 
typically). Here the value of $\Delta n_R$ is very small \cite{d1, d2},  
increases above the threshold and then decreases at large $\omega $ as
\be
\Delta n_R^h(\omega )\approx \frac {a}{\omega },     \label{dhig}
\ee
where $a\approx 2\cdot 10^{-3}$ GeV  at $\rho \approx 0.1$ as follows 
from experiment and Eq. (\ref{ren}). The dispersion relations show 
that real parts of other hadron  amplitudes also become positive. The 
index $h$ refers to high energy hadrons.

One can speculate that approximately the same estimates should be valid for
gluons as carriers of strong forces if this feature is common for all
hadronic reactions. Then they can be used to get some insight into the 
characteristics of the collective conical flow of parton jets initiated 
by a primary parton piercing through a nuclear slab. In principle, one should 
consider the index of refraction for initial parton (quark or gluon) as well.
However, since its energy is extremely high, then $\Delta n$ becomes very
close to 0 according to (\ref{dhig}). 

The nuclear index of refraction was recently discussed in connection with
Cherenkov gluons in \cite{mwa, kmwa}.

Partons able to emit gluons with energies above the mentioned threshold and
those which can emit only "resonance" gluons are treated separately below.

\subsection{Very high energy jets}

First, let consider the forward moving jet produced by an initial parton.
In the target rest system it has very high energy if cosmic rays or
Tevatron, RHIC, LHC energies\footnote{Note that LHC energies correspond to 
10$^{17}$ eV in the target rest system while the threshold energy $\omega _{th}$ 
is about 10$^{11}$ eV.} are considered. The emitted gluons can be
also very energetic. For would be infinite hadronic medium  and gluons with
energies above the threshold one gets from (\ref{cos}) and (\ref{dhig}) the 
cone angle
\be
\theta_{\infty }=\sqrt {2\Delta n_R^h(\omega )}=\sqrt {\frac {3m_{\pi }^3}
{4\pi \omega }\sigma (\omega)\rho (\omega )}.   \label{tinf}
\ee

However, the size of hadronic targets is very small $l\approx A^{1/3}/m_{\pi }$
where $A$ is the atomic number of a nucleus. The color currents appear
only inside the target during the collision. Partons are confined and color
neutralized outside it. If, following I. Tamm \cite{tamm}, one considers
emission by a current acting only at the finite length $l$ and separates
its part proportional to $\Delta n_R$ that is typical for Cherenkov radiation,
the following distribution is obtained \cite{d2}
\be
\frac{1}{\sigma }\frac {d\sigma }{d\omega dz}=\frac {4\alpha_sC_Fl\Delta n_R^h}
{\pi }f(z),   \label{sig}
\ee
where
\be
f(z)=\frac {\sin z}{z}\left[\frac {\sin z}{z}-\cos z\right],   \label{fz}
\ee
\be
z=\frac {\omega l\theta^2}{4}=\frac {p_tl\theta }{4}< \pi.   \label{z}
\ee
The function $f(z)$ has a maximum at
\be
\theta=\sqrt {\frac {2\pi }{ \omega l}}  \label{tmax}
\ee
and the width
\be
\Delta (\ln \theta )< 0.4\; -\; 0.5.   \label{wid}
\ee
The conical flow appears at smaller angle (\ref{tmax}) compared to 
$\theta _{\infty}$  (\ref{tinf}). But this angle is much larger than the
bremsstrahlung angles $\theta _{bs}\propto 1/E$ ($E$ is the energy of the 
initial parton).

The transverse momenta are large.
\be
p_t=\sqrt {\frac {2\pi \omega}{ l}}.  \label{pt}
\ee
They are in GeV-region for  $\omega >\omega _{th}$.

The crucial parameter which differs the finite length emission from the
infinite one is $\omega l \Delta n_R/2$. The formula (\ref{tinf}) is valid 
for  
\be
\omega l \Delta n_R/2\gg 1.   \label{gg}
\ee
For estimates of $\Delta n_R$  given by Eq. (\ref{dhig}) one gets the opposite 
condition $\omega l \Delta n_R/2\ll 1$ for all realistic $l$. That is why
the condition (\ref{tmax}) arises.

Let us notice that the imaginary part of the index of refraction describes
the attenuation of the created flow. For a plane wave e$^{ikr}$ it is given 
by a factor $\exp [-\omega r{\rm Im} n(\omega )]=\exp[-3m_{\pi }^3\sigma r/8\pi ]$
which damps it down at distances by the order of magnitude larger than 1 Fm,
i.e. our treatment should be valid practically for all nuclei.

The primary parton creates a forward moving jet which would correspond 
in the above classical model to the term not vanishing at $\Delta n_R
\rightarrow 0$ and omitted in the calculations. It can not be treated by 
this model but QCD works quite well in describing such jets. Namely this 
jet initiates the conical flow. The microscopic mechanism of this effect
is the collective excitation and polarization of colored partons in the 
medium by a forward moving parent with subsequent coherent directed 
emission of color radiation.

The cone angle for high energy gluons in the rest system of target (\ref{tmax}) 
is rather small
($\theta \ll 1$) because $\omega \gg l^{-1}$. However, being transformed 
to c.m.s. of colliding nuclei, where RHIC experiments are done, the angles 
become large, and, moreover, the 
cones can exchange by hemispheres. This is the general relativistic effect
of "medium" motion.
The approximate estimates given in \cite{d2} indicate that for 
$pp$-collisions this effect could be noticed at rather large c.m.s. 
angles about 60$^{\rm o}$ - 70$^{\rm o}$, i.e. at quite small pseudorapidities
$\vert \eta _c\vert \approx 0.3 - 0.5$. To get
more precise estimates, some additional assumptions are necessary.

According to (\ref{tmax}), this angle is smaller for nucleus collisions.
For two nuclei colliding, two humps in the pseudorapidity distribution
must appear. It follows from Eq. (\ref{tmax}) that the cone angle 
is proportional to $l^{-1/2}$, i.e. to $A^{-1/6}$. It becomes smaller for 
larger nuclei. In the pseudorapidity plot this corresponds to the shift
of humps by $\frac {1}{6}\ln A$ that is close to 1 for largest nuclei 
compared to  nucleon. Since any of the colliding nuclei can be treated 
as a target or as a bunch of partons, two cones in the opposite hemispheres 
in c.m.s. may be observed. They are symmetrical for identical nuclei. 
Especially interesting is the difference between predictions for different 
nuclei. If a light nucleus collides with a heavy one, forward and backward
cones have different angles according to the above comment. For larger $l$ this 
dependence becomes slower and $\theta $ approaches $\theta _{\infty}$ so
that no such effect exists if $\Delta n_R$ or $l$ are so large that the 
condition (\ref{gg}) is satisfied.

The important 
corollary of Eq. (\ref{dhig}) is that a high energy gluon can also emit
Cherenkov gluons due to self-interaction of gluons. The phase velocity 
of emitted gluon must be smaller than the phase velocity of initial one.
This is fulfilled if $\Delta n_R^h(\omega )$ decreases with $\omega $
(see (\ref{dhig})). In this case secondary Cherenkov rings around the 
direction of emitted Cherenkov gluon can be detected, in principle.

Very high energy partons can also emit rather soft Cherenkov gluons 
corresponding to resonance regions for hadrons. The emission angle in the 
target rest system is rather large for forward moving jets in accordance  with
(\ref{cos}) and (\ref{del}). One can try to detect this effect only in the 
fixed target experiments. It results in two peaks in the deep forward and
backward fragmentation regions of the pseudorapidity distribution. These peaks
are unobservable in collider experiments because particles are captured inside
the accelerator ring. For "resonance" gluons and large nuclei, the condition
(\ref{gg}) can be satisfied. Then the emission angle is defined by
\be
\cos \theta _r\approx \frac {1}{1+\Delta n_R^r}    \label{thr}
\ee
and does not depend on $l$. 

\subsection{Moderate energy jets}

Instead of producing the forward moving energetic jet, the primary parton 
can in some cases create secondary jets with lower energies. This can be 
either the result of its hard high-$p_t$ scattering or of production
of a new aside moving jet similar to the gluon jet in three-jet events
in $e^+e^-$-annihilation. Kinematics of these two processes is 
different, in general. If the energy of a secondary jet is below the
high energy threshold $\omega _{th}$ for positiveness of $\Delta n_R(\omega)$, 
then only
low energy gluons relevant for production of hadron resonances can be emitted. 
The above estimates of $\Delta n_R^r$ show that the parameter
$\omega l \Delta n_R^r/2$ is of the order of 1. The cone angle becomes 
large even in the target rest system (see (\ref{thr})). The attenuation of
such gluons is quite strong because Im$\Delta n^r$ is of the order of 1. They
can produce resonances. Therefore, one can speculate that such moderate energy 
jets would be accompanied by ring regions with some excess of resonances forming
a kind of "resonance gas" for a short time. This effect is pronounced 
only in the special narrow energy bands of corresponding wings of 
resonances. The attempts to define these bands were done in \cite{zlom, isto, 
isar} without referring to the parton structure of particles.
For large values of $\Delta n_R^r$ the transition radiation induced by 
the target edges can also become important.

\subsection{Conclusion on Cherenkov gluons}

Thus, one can expect to observe two (high and low energy) components of 
Cherenkov gluons if Eq. (\ref{nom}) is applicable in QCD and the data on 
real parts of hadronic amplitudes are used. 

High energy Cherenkov gluons can be emitted by the primary impinging parton at 
small polar angles in the target rest system given by Eq. (\ref{tmax}). Each 
gluon would produce a jet. These jets should form the ring with "jetty"
substructure (spots) in the azimuthal plane perpendicular to the direction of 
the primary parton. At very high energies the number of Cherenkov gluons 
can be large and they form a ring in a single event. If
one or few gluons are emitted in an event, the rings can be detected only
in high statistics experiments as peaks of the pseudorapidity distribution. 
The emission angle becomes quite large in 
the system where the target moves towards the impinging parton (collider
experiments). The transverse momenta of particles inside the ring can be 
somewhat enhanced.

For low energy Cherenkov gluons corresponding to resonance regions, the emission
angle is large even in the target rest system, i.e., in the fixed target
experiments. In the collider (c.m.s.) system
it would correspond to extremely large polar angles for forward moving initial
parton and effect is not observable. The effect can be observed only for 
transversally moving partons or, may be, in collisions of protons with heavy 
nuclei. The enhancement of resonance production can influence the $\pi /p$
ratio inside the rings compared to outer regions. Almost monochromatic pions
would dominate, if resonance motion and confinement effects do not spoil it 
for such low energies.

Small values of $\Delta n_R >0$ in the high 
energy region and narrow energy bands  in which $\Delta n_R >0$ at low 
energies imply that the probability of the process is not very high and
it should be carefully chosen from large background due to more ordinary
processes.

\section{Mach waves}

Any quark-gluon jet created by a high energy parton can become a source 
of another collective conical flow which has macroscopic analogue known
as Mach shock waves. Since the jet propagates with the speed of light it 
can be considered as a supersonically moving body. Therefore the large 
perturbations of matter are created and give rise to shock waves. They  
would induce particle radiation at Mach angle $\theta _M$ with the jet 
direction in the target rest system
\be
\cos \theta _M=\frac {c_s}{c}.    \label{mach}
\ee
The cone of the front of the shock wave has the angle $\theta _f=
\frac {\pi }{2}-\theta _M$.
Thus, at the microscopic level, the cone of particles should be produced.
Now, everything depends on how one treats the hadronic medium \cite{glas,
baum, sche, khod, shmg, chgr, rsgr, shur, stoc, mrup} and whether sound waves
can propagate in it. Unfortunately, our knowledge of this medium is still 
not complete enough to answer these questions. In various models and at
different conditions, its states range from quite dilute cold nuclear 
matter and nuclear Fermi liquid to quark-gluon condensate (CGC), 
quark-gluon plasma (QGP), strongly interacting quark-gluon plasma (sQGP),
"resonance gas" and dense quark-gluon liquid (QGL). The Mach angle does 
not depend on jet energy but depends on the state of matter. For weak waves
in the mixed phase state the speed of sound is negligibly small $c_s\approx 0$.
The lattice 
calculations \cite{gava} show that the speed of sound can be still rather small
$c_s^2\approx 0.15c^2$. For "resonance gas" it was estimated \cite{shur}
equal to $c_s^2\approx 0.2c^2$. For ultrarelativistic equation of state
it should be $c_s^2= \frac {1}{3}c^2$. For sQGP the value 
$c_s^2= \frac {2}{3}c^2$ may be chosen. For plasma with strong vector
(gluon!) interaction the shock waves are strong and have larger speeds
$c_s\approx c$, i.e. the situation approaches the conditions for high 
energy Cherenkov gluons.

The cone angles are quite large ($ 60^{\rm o}$ if $c_s=0.5c$) in the rest 
system of the target. Mach cones produced by forward jets initiated by primary 
partons are reversed to large pseudorapidities in the backward hemisphere, 
when transformed to c.m.s. of colliding nuclei. For RHIC energies, they escape 
in accelerator pipes unobserved (see Section 5). This effect, however, can be 
detected in fixed target experiments or for 
transversally moving jets (in c.m.s.) similarly to low energy Cherenkov gluons.

\section{Experimental data}

The most important feature of both effects is the cone of particles (or 
subjets) at the definite angle to the direction of propagation of initiating 
it parton (or jet). If this direction is chosen as $z$-axis, then there 
should be maximum
in pseudorapidity ($\eta =-\ln \tan \theta /2$) distribution of particles 
for a corresponding event or a set of events. Two colliding nuclei must
give rise to two peaks. Namely such maxima observed 
in the cosmic rays event at energy $10^{16}$ eV \cite{addk} (Fig. 1) 
initiated the idea about Cherenkov gluons \cite{d1}. When the two-dimensional
distribution of particles was considered in the azimuthal plane (called 
as target diagram in cosmic rays experiments), this event revealed two
(forward and backward in c.m.s.) densely populated ring-like regions
within narrow interval of polar angles but widely distributed in 
azimuthal angles. Therefore such events were called as ring-like events.

Approximately at the same time the similar event with two peaks was 
observed at $10^{13}$ eV \cite{arat}. Some events with one peak were 
shown even earlier \cite{alex, masl}. Individual events with a prominent
ring-like structure were also found at accelerator energies \cite{maru, 
kit}. Especially impressive is the event \cite{kit} with peak in
the pseudorapidity distribution $dn/d\eta =100$ within narrow interval
$\delta \eta =0.1$ that 60 times exceeds the average density (Fig. 2).
As usually, for individual events, it is unclear whether they reveal some 
dynamical mechanism or are rare statistical fluctuations even though it 
has been estimated that probability of these fluctuations is extremely 
low.

The experimental data on high multiplicity hadron collisions with high
statistics at energies of hundreds GeV became available only in 1980s. 
It would be naive to expect to observe the peaks at these energies 
immediately because the background from ordinary processes is huge.
One had to select events by some criteria which would favor observation
of subjets, i.e. groups of particles quite dense on the pseudorapidity
plot and well separated from other groups. One can hope that the effect 
is more clearly seen with such a choice and the background is less strong.
This program was applied to $pp$-interactions in fixed target 
experiments at 205 and 360 GeV 
\cite{dlln} and later to $\pi p$ and $Kp$ interactions \cite{agab}.
The pseudorapidity distributions of the centers of dense isolated
groups (which were interpreted as jets) possess the peaks at 
$ \eta _c=\pm 0.3$, i.e. at $\theta _{c.m.s.}\approx 70^{\rm o}$ and
110$^{\rm o}$ (Fig. 3). No such effect is seen in Fritiof 1.6 events.

Apart from considering the distribution of centers of dense isolated groups
one can analyze the azimuthal structure \cite{adam, ad2, voka} plotting
the sum of squared differences of azimuthal angles of two neighboring 
particles in the investigated groups. It must be different for purely 
stochastic particle production, for jet-like mechanism and for ring-like
events. Observed distributions also favor presence of ring-like
events in interactions Pb-emulsion (Ag, Br) at 158 GeV and Au-emulsion 
at 11.6 GeV. Events with this azimuthal parameter favoring ring-like
structure show pronounced two-hump property of the pseudorapidity
distribution.

A new method of pattern recognition in event-by-event analysis of 
two-dimensional correlations 
in individual high multiplicity events using wavelets \cite{dine} was
applied to the shown above cosmic rays event \cite{adk} and to fixed target
PbPb interactions at 158 GeV \cite{dikk}. The ring-like structure in the 
target diagram of some events was found with radii corresponding to
the maxima in pseudorapidity distribution. Fig. 4 demonstrates how 
correlations at some scale are distributed in the azimuthal plane of 
one of PbPb events at 158 GeV. The dark regions forming jetty rings
correspond to correlated groups of particles (jets).

Some low-energy data on AA-interactions showed similar surprising 
irregularities as well \cite{elna, gogi}. In principle, the positive
values of $\Delta n_R^r$ in one of the resonance wings could be blaimed 
for them.

A new approach to search for ring-like events at RHIC and LHC energies
was recently proposed in \cite{dste}. HIJING model has been used to get 
inclusive pseudorapidity distribution for 3500 central ($b$=0) collisions
(AuAu at $\sqrt s =200$A GeV and PbPb at $\sqrt s =5500$A GeV). Then 
the spikes in 
individual HIJING events exceeding this distribution by more than one or two 
standard deviations are separated and the distribution of their
centers is plotted. They are considered as purely statistical 
fluctuations or jets produced at hard scattering. Therefore this 
distribution is rather smooth.
If experimental data show some peaks at definite
pseudorapidity values over this background, this can be an indication 
on a new collective effect not considered in HIJING.

The above data concerned mostly either individual cosmic rays events or 
low-statistics nucleus-emulsion interactions. The high statistics data 
were available only for hadron-hadron interactions at relatively low 
energies and asked for some special selection in the primary data as 
discussed above. Gluons emitted by forward moving jets were 
searched for. The recent high-statistics  data on AA-interactions 
from RHIC allow to search for new peculiar features using a special 
trigger \cite{wang}. Instead of evolution of a primary forward moving 
jet, the process of creation of two oppositely moving jets by a primary parton 
was chosen. These moderate energy jets originate at some point inside the target
nucleus and move transversally to the initial parton, i.e. in the 
azimuthal plane. This shows that they are produced in high-$p_t$ 
head-on collisions of two partons belonging to the colliding nuclei. 
Surely, the choice of this quite rare process with 
high transferred momentum drastically reduces the statistics. It is 
compensated by a good trigger provided by the jet which passes only the 
thin layer of matter near the edge of the nucleus. Therefore one can 
select these events and study
the effects induced by its companion jet piercing through the larger part
of the nucleus in the opposite direction. The trigger jet produces the
particle flow near $\phi =0$ and the companion jet - near $\phi =\pi $.
These peaks are clearly seen in $pp$-collisions. However, in AuAu 
collisions the distribution near $\phi =\pi $ is double-peaked one
with a minimum at $\phi = \pi $ (see Fig. 5). 
This can be explained as a result of 
cone radiation. Since the companion jet moves in the same azimuthal
plane where the observation is done, the cone is cut by the plane which 
contains its vertical. Therefore two peaks along the two cut sides of 
the cone are detected. They add to the usual companion jet radiation
which moves at $\phi =\pi $ and can be considered here as background.
If the cone were cut by the plane perpendicular to its
vertical, i.e. to the direction of propagation of the companion jet, 
as it was done for primary jets, one would see the 
ring-like structure of these events. It is interesting to perform 
measurements in the plane perpendicular to the cone vertical and to see 
directly the ring-like structure.

This effect is interpreted in \cite{shur} as Mach waves moving with the 
speed $c_s=0.33c$. It is claimed in \cite{shur} that this value of $c_s$
can be obtained for RHIC as a time-weighted average of several stages
of nuclear matter: QGP, mixed phase and "resonance gas".
The c.m.s. system of RHIC coincides with the
"laboratory system" of the transversally moving companion jet. The angle
$\theta _M\approx 70^{\rm o}$ is quite large. 

Let us note that the typical energies of the companion jets are about 
5-20 GeV which are much below the energy of a primary parton at RHIC if 
considered in the target rest frame. They are 
also  below those energies where $\Delta n_R$ becomes positive at high 
energies in hadron collisions. However, namely for resonances $\Delta
n_R^r$ becomes positive and rather large according to (\ref{del}).
This would provide quite large angles for Cherenkov gluons in this 
energy range as well. The angle 70$^{\rm o}$ corresponds to 
$\Delta n_R^r=2$ according to Eq. (\ref{thr}).

One is tempted to conclude that spikes of the
pseudorapidity distributions at large c.m.s. angles and peaks for companion
jets observed at RHIC are initiated by jets with high and moderate energies,
correspondingly, if treated as Cherenkov gluons.

Any jet created with the same energy in fixed target experiments should give 
rise to the similar effect. It is surprising that the peaks at large angles
in the target system
were not observed earlier for forward moving jets in those fixed target 
experiments, where the conditions similar to RHIC (jets with energies about 
10 GeV inside a nucleus at rest) were achieved.

\section{Specifics of collective hadronic events}

Theoretically, Cherenkov and Mach waves are different because they are
transverse and longitudinal excitations, correspondingly. However, it 
is not easy to reveal this feature in experiment. Their common feature 
is the cone of radiation. It must result in two humps of the 
one-dimensional pseudorapidity distribution and in ring-like structure
of two-dimensional plots. They have been observed. Cherenkov and Mach 
waves could differ by their most prominent characteristics, the emission
angle (\ref{cos}). From above discussion one sees that the sound
velocity can range from very low values for weak waves to $c$ for strong waves.
The phase velocity of gluons would be close to $c$ at high energies and
several times less at low energies if above estimates applied. Thus
there is strong overlap  in these numbers and further properties must
be studied.

Each high energy gluon would produce a jet. If the
number of emitted Cherenkov gluons in a single event is not large, this
event would not have the ring-like structure but jets preferring the angle 
(\ref{cos}). For Mach waves, the ring-like structure must appear in any 
individual event until we treat them in a classical way and do not speak about
phonons, plasmons (strongly interacting!?) etc. Specific dependence of 
the emission angle on atomic number could differ these two mechanisms. 

The enhanced transverse momenta could be a signature for high energy 
Cherenkov gluons. The latest D0 data \cite{wob} on jet transverse momenta 
up to 510 GeV in $pp$-interactions at $\sqrt s =1.96$ TeV have been 
successfully fitted by QCD. Thus the enhancement should be a purely nuclear 
effect. This conclusion would agree with RHIC data but indirectly differs
from results of \cite{dlln}.

Gluons differ from photons by self-coupling.
Therefore, a high energy Cherenkov gluon can emit its own secondary gluon.
If well separated  as 3-jet events in $e^+e^-$-annihilation, both can 
give rise to ring-like  structures in the planes perpendicular to their
directions. (Two gluons producing smoke-rings!) However, the intensity of 
this secondary effect must be very low if the initial effect is not 
strong enough and would require the special trigger. 

Low energy 
Cherenkov gluons can be efficient  in producing resonances. This would
change the pion momentum spectrum and $\pi /p$ ratio in the rings more 
densely populated with resonances compared to outside regions.

Eq. (\ref{cos}) defines the emission angle in the target system.
The forward moving jet meets the backward moving flow of partons in c.m.s..
The medium motion relative to any jet can be easily accounted and
characteristics in different coordinate systems calculated. In particular,
if the forward moving parton produces Mach waves at $c_s=0.33c$ which generate
massless partons, the emission angle in the RHIC system differs from 
180$^{\rm o}$ less than by 1$^{\rm o}$:
\be
\theta _{c.m.s.}=\pi-\frac {1}{\gamma }\left( \frac {c+c_s}{c-c_s}\right)^{1/2}
\approx 180^{\rm o}-0.8^{\rm o}.   \label{tcms}
\ee
The difference from $\pi $ is even less for massive particles. This angle
is of the order of bremsstrahlung angles in the deep fragmentation region.
Thus, the effect can not be observed. (It flies in the tube!)

The situation is not completely clear for aside moving jets. One can find 
out from the shift of the cone whether the medium motion is important.
The symmetrical positions of peaks at RHIC energies favor the conclusion 
that the medium is as a whole at rest for transverse jets. However, there 
exists more exotic possibility that the matter is not equilibrated and 
partons still remember their initial directions with 
two flows moving in opposite directions like in the superfluid liquid
even though the center of mass is at rest.
The cone angle will be changed. In statistical physics, this state of matter
asks for the second virial coefficient becoming negative and particles paired.
In QCD, the analogous effect of negative cumulant moments 
\cite{dre3, 41, dgar} can be related to the similar phenomenon with
many-parton clustering in place of two-particle pairing.

\section{Conclusion}

There are some indications from experiment on coherent collective
effects in hadronic matter which result in the so-called ring-like
events. They can be explained as cones formed either by Cherenkov gluons
or by Mach shock waves. Cherenkov and Mach waves have similar origin but 
correspond to propagation of transverse and longitudinal excitations in 
a medium, correspondingly. High-multiplicity central nuclei collisions 
are preferred for their search because of larger number of participating
partons even though the background increases also.

Two peaks of the pseudorapidity distribution positioned in accordance 
with Eq. (\ref{cos}) provide most important 
signature of ring-like events in one-dimensional plots. Two-dimensional
plots in the plane perpendicular to the motion of initiating jet 
must have ring-like structure. More detailed 
characteristics of these events are necessary to make definite choice.
In particular, energy and A-dependences of this effect as well as
resonance contents and transverse momentum behavior must be studied. 
More direct proof of the ring-like
structure in high statistics experiments is necessary, e.g., the
analysis of RHIC events in the plane perpendicular to the companion
jet direction, i.e. to the cone vertical. The observation of ring-like
structure induced by secondary gluons would be very important. One can
hope to get the value of the nuclear index of refraction and/or the
sound velocity in nuclear matter from the cone angle. This can lead us 
to the proper equation of state of the nuclear matter.

Anyway, there is no doubt that study of these events will provide us 
with deeper knowledge of properties of quark matter at extreme energies.\\
{\bf Acknowledgements}\\
I am grateful to V.A. Nechitailo and E.K. Sarkisyan for help with Figures.
This work has been supported in part by the RFBR grants 03-02-16134, 
04-02-16445-a, NSH-1936.2003.2.\\

\newpage

\begin{figure}
\center{\includegraphics[bb=20 620 540 820, width=8.cm]{Fig1.ps}}
\vspace*{3.cm}
{\caption{
The pseudorapidity distribution of produced particles in the 
stratospheric event at 10$^{16}$ eV \cite{addk} has two pronounced peaks.}
}
\end{figure}

\newpage

\begin{figure}
\center{\includegraphics[bb=50 340 540 800, width=5.cm]{Fig2.ps}}
\vspace*{6.cm}
{\caption{
The pseudorapidity distribution of NA22-event \cite{kit} with
extremely high peak.}
}
\end{figure}

\newpage

\begin{figure}
\center{\includegraphics[bb=60 120 580 820, height=1.cm]
{Fig3.ps}}
\vspace*{21.cm}
{\caption{
The pseudorapidity distribution of the centers of dense isolated
groups in $pp$-interactions at 360 GeV \cite{dlln} shows some excess over
background at $\vert \eta _c\vert \approx 0.3$.}
}
\end{figure}

\vspace*{5.cm}
\newpage

\begin{figure}
\center{\includegraphics[bb=47 210 530 800, height=13.cm]
{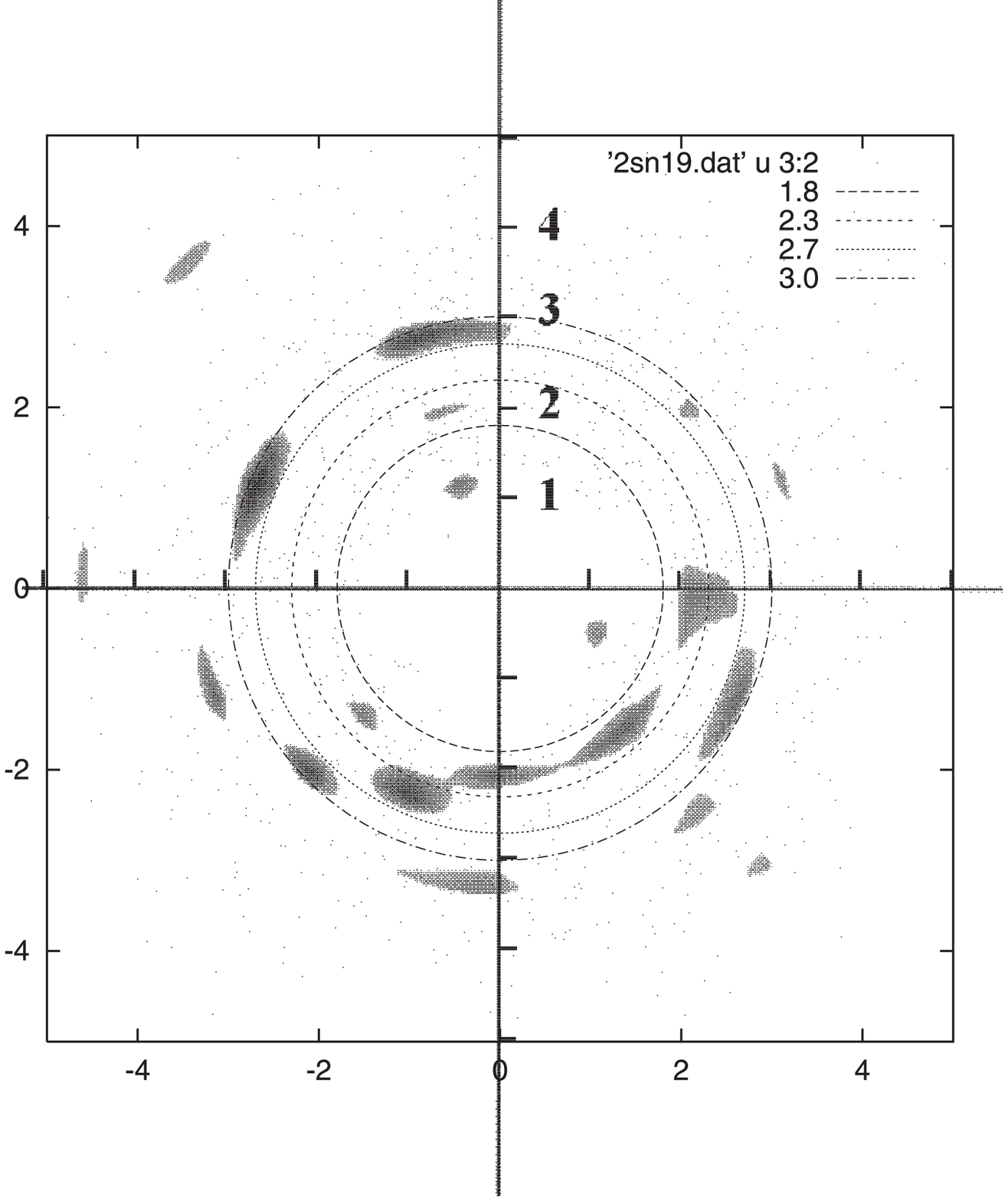}}
\vspace*{2.cm}
{\caption{
The two-dimensional $(\eta -\phi $) plot of  the wavelet 
coefficients at scale $j=5$ obtained from analysis of PbPb event at 
158 GeV \cite{dikk}. Dark regions denote large values of wavelet 
coefficients, i.e. strongly correlated groups of particles.
The ring regions $1.8<\eta <2.3$ and $2.7<\eta <3.0$ correspond to peaks
of the pseudorapidity distribution $\eta =2.5$ is equivalent to c.m.s.
angle $\pi /2$. The dots indicate particle positions. The empty space
near the center is due to the limited acceptance of the detector.}
}
\end{figure}

\newpage

\begin{figure}
\center{\includegraphics[bb=40 120 572 754, width=4.cm]{Fig5.ps}}
\vspace*{18.cm}
{\caption{
The $\phi $-distribution of particles produced by trigger and 
companion jets at RHIC \cite{wang} shows two peaks in $pp$ and three 
peaks in AuAu-collisions.
}
}
\end{figure}


\newpage

\begin{thebibliography}{99}
\bibitem{d1}
 I.M. Dremin, Pisma v ZhETF  30 (1979) 152; JETP Lett. 30 (1979) 140.
\bibitem{d2}     
 I.M. Dremin, Yad. Fiz. 33 (1981) 1357; Sov. J. Nucl. Phys. 33 (1981) 726.
\bibitem{addk}
A.V. Apanasenko, N.A. Dobrotin, I.M. Dremin, K.A. Kotelnikov,
Pisma v ZhETF  30 (1979) 157; JETP Lett. 30 (1979) 145.
\bibitem{alex}
K.I. Alexeeva et al, Izv. AN SSSR 26 (1962) 572; J. Phys. Soc. Japan,
1962, 17, A-II.
\bibitem{masl}
N.V. Maslennikova et al, Izv. AN SSSR 36 (1972) 1696.
\bibitem{arat}
N. Arata, Nuovo Cim. 43A (1978) 455.
\bibitem{maru}
E.A. Marutyan et al, Yad. Fiz. 29 (1979) 1566.
\bibitem{kit}
M. Adamus et al (NA22), Phys. Lett. B185 (1987) 200.
\bibitem{dlln}
I.M. Dremin et al, Mod. Phys. Lett. A5 (1990) 1743; Yad. Fiz. 52 (1990)
862, Sov. J. Nucl. Phys. 52 (1990) 840. 
\bibitem{adam}
M.I. Adamovich et al, J. Phys. C21 (1993) 2035.
\bibitem{elna}
A. El-Naghy et al, Phys. Lett. B299 (1993) 370.
\bibitem{agab}
N.M. Agababyan et al (NA22), Phys. Lett. B389 (1996) 397.
\bibitem{adk}
N.M. Astafyeva, I.M. Dremin, K.A. Kotelnikov, Mod. Phys. Lett. A12 
(1997) 1185.
\bibitem{swan}
S. Wang et al, Phys. Lett. B427 (1998) 385.
\bibitem{ad2}
M.I. Adamovich et al, Eur. Phys. J. A5 (1999) 429.
\bibitem{gogi}
G.L. Gogiberidze et al, Phys. Lett. B430 (1998) 368, B471 (1999) 257; Yad.  Fiz. 
64 (2001) 147, 
Phys. Atom Nucl. 64 (2001) 143;  Nucl Phys. Proc. Suppl. 92 (2001) 75.
\bibitem{dikk}
I.M. Dremin et al, Phys. Lett. B499 (2001) 97.
\bibitem{voka}
S. Vokal et al, hep-ex/0412017; 0501025.
\bibitem{wada}
W. Wada, Phys. Rev. 75 (1949) 981.
\bibitem{igur}
D.D. Ivanenko, V.A. Gurgenidze, DAN SSSR, 67 (1949)  997.
\bibitem{bind}
D.I. Blokhintsev, V.L. Indenbom, ZhETF 20 (1950) 1123.
\bibitem{yeku}
G. Yekutieli, Nuovo Cim. 13 (1959) 446, 1306.
\bibitem{ceri}
W. Czyz, T. Ericson, S.L. Glashow, Nucl. Phys. 13 (1959) 516.
\bibitem{cgla}
W. Czyz, S.L. Glashow, Nucl. Phys. 20 (1960) 309.
\bibitem{smrz}
P. Smrz, Nucl. Phys. 35 (1962) 165.
\bibitem{inic}
D.B. Ion, F.G. Nichitiu, Nucl. Phys. B29 (1971) 547.
\bibitem{zlom}
D.F. Zaretskii, V.V. Lomonosov, Sov. J. Nucl. Phys. 26 (1977) 639.
\bibitem{mwa}
A. Majumder and Xin-Nian Wang, nucl-th/0507062.
\bibitem{kmwa}
V. Koch, A. Majumder and Xin-Nian Wang, nucl-th/0507063.
\bibitem{tamm}
 I.E. Tamm, J. Phys. USSR  1 (1939) 439.
\bibitem{isto}
D.B. Ion, W. Stocker, Phys. Lett. B273 (1991) 20, B346 (1995) 172; 
Phys. Rev. C52 (1995) 3332.
\bibitem{isar}
D.B. Ion, E.K. Sarkisyan, hep-ph/0302114.
\bibitem{glas}
A.E. Glassgold, W. Heckrotte, K.M. Watson, Ann. Phys. 6 (1959) 1.
\bibitem{baum}
H.G. Baumgardt et al, Z. Phys. A273 (1975) 359.
\bibitem{sche}
W. Sch\"afer et al, Z. Phys. A288 (1978) 349. 
\bibitem{khod}
V.A. Khodel et al, Phys. Lett. B90 (1980) 37.
\bibitem{shmg}
H. St\"ocker et al, Prog. Part. Nucl. Phys. 4 (1980) 133.
\bibitem{chgr}
C.F. Chapline, A. Granik, Nucl. Phys. A459 (1986) 68; A511 (1990) 747.
\bibitem{rsgr}
D.H. Rischke, H. St\"ocker, W. Greiner, Phys. Rev. D42 (1990) 2283.
\bibitem{shur}
J. Casalderrey-Solana, E.V. Shuryak, D. Teaney, hep-ph/0411315.
\bibitem{stoc}
H. St\"ocker, Nucl. Phys. A750 (2005) 121.
\bibitem{mrup}
J. Ruppert and B. Muller, Phys. Lett. B618 (2005) 123.
\bibitem{gava}
R. Gavai, S. Gupta, S. Mukherjee, Phys. Rev. D71 (2005) 074013; hep-lat/0506015.
\bibitem{dine}
I.M. Dremin, O.V. Ivanov, V.A. Nechitailo, UFN 171 (2001) 465, Physics-Uspekhi
44 (2001) 447.
\bibitem{dste}
I.M. Dremin, L.I. Sarycheva, K.Yu. Teplov, talk at QM2005, Budapest.
\bibitem{wang}
F. Wang (STAR), J. Phys. G30 (2004) 1299.
\bibitem{wob}
M. Wobisch (D0), hep-ex/0411025.
\bibitem{dre3}
I.M. Dremin, Phys. Lett. B 313 (1993) 209.   
\bibitem{41}
I.M. Dremin, V.A. Nechitailo, Mod. Phys. Lett. A9 (1994) 1471;
JETP Lett. 58 (1993) 881.
\bibitem{dgar}
I.M. Dremin, J.W. Gary, Phys. Rep. 349 (2001) 301.

\end{thebibliography}
\end{document}